\documentclass[aps,ams,prl,showpacs,twocolumn,floatfix,superscriptaddress,reprint]{revtex4} 
\usepackage{amssymb} 
\usepackage[dvips]{color} 

\usepackage{graphicx} 

\newcommand{\p}{\partial}

\newcommand \be {\begin{equation}}
\newcommand \ee {\end{equation}}
\newcommand \bea {\begin{eqnarray}}
\newcommand \eea {\end{eqnarray}}

\newcommand \w {\mathbf{w}}

\begin{document}

\title{Continuous theory of active matter systems with metric-free interactions}

\author{Anton Peshkov}
\affiliation{Service de Physique de l'Etat
  Condens\'e,~CEA-Saclay,~91191~Gif-sur-Yvette,~France}
\affiliation{LPTMC, CNRS-UMR 7600, Universit\'e Pierre et Marie Curie, 75252 Paris, France}
\author{Sandrine Ngo}
\affiliation{Service de Physique de l'Etat Condens\'e,~CEA-Saclay,~91191~Gif-sur-Yvette,~France}
\author{Eric Bertin}
\affiliation{Universit\'e de Lyon, Laboratoire de Physique, ENS Lyon, CNRS, 46 All\'ee d'Italie, 69007 Lyon, France}
\author{Hugues Chat\'e}
\affiliation{Service de Physique de l'Etat Condens\'e,~CEA-Saclay,~91191~Gif-sur-Yvette,~France}
\author{Francesco Ginelli}
\affiliation{Istituto dei Sistemi Complessi, CNR, via dei Taurini 19, I-00185 Roma, Italy}
\affiliation{Institute for Complex Systems and Mathematical Biology, King's College, University of Aberdeen, Aberdeen AB24 3UE, United Kingdom}

\date{\today} 

\begin{abstract}
We derive a hydrodynamic description of metric-free active matter: 
starting from self-propelled particles aligning with neighbors 
defined by ``topological'' rules, not metric zones, 
---a situation advocated recently to be relevant for bird flocks,
fish schools, and crowds---
we use a kinetic approach to obtain well-controlled nonlinear 
field equations. We show that
the density-independent collision rate per particle 
characteristic of topological interactions suppresses the linear instability of the
homogeneous ordered phase and the nonlinear density segregation 
generically present near threshold in metric models, 
in agreement with microscopic simulations. 
\end{abstract}

\maketitle 

Collective motion is a central theme in the emerging field of 
active matter studies \cite{SR-REVIEW}. 
For physicists, the interest largely lies in the non-trivial cases
where the emergence of collective motion can be seen as an instance of 
spontaneous symmetry-breaking out of equilibrium: without leaders, 
guiding external fields, or confinement by boundaries, 
large groups inside which an ``individual'' can only perceive
local neighbors are able to move coherently. 
After this was realized in the seminal papers of Vicsek {\it et al.}
\cite{Vicsek95} and 
Toner and Tu \cite{TT}, much progress has been recorded in the physics community
\cite{SR-REVIEW,Reviews}, alongside continuing 
modeling work in ethology and biology \cite{ETHO-BIO,Swinney2010}. 

Most models consist of self-propelled particles 
interacting with neighbors defined to be those particles within some
finite distance \cite{Parrish}. 
Among those ``metric models'', that introduced by Vicsek  {\it et al.} \cite{Vicsek95}
is arguably the simplest: in the presence of noise,
point particles move at constant speed, aligning ferromagnetically with others 
currently within unit distance. 
The study of the Vicsek model has revealed rather unexpected behavior. Of particular 
importance in the following is 
the emergence of phase segregation, under the form of high-density 
high-order traveling bands \cite{Gregoire2004},
in a large part of the orientationally-ordered phase bordering the 
onset of collective motion,
relegating further the spatially-homogeneous fluctuating phase treated by 
Toner and Tu.
Similar observations of density segregation were made for
important variants of the Vicsek model, such as polar particles with
nematic alignment \cite{Ginelli2010}
(self-propelled rods) or the active nematics model of 
\cite{Chate2006, Ramaswamy2}.
The genericity of these observations has been confirmed, 
in the Vicsek case, 
by the derivation and analysis of continuous field equations
\cite{bertin, Ihle} (see also \cite{MCM, MCM2}). 
It was shown in particular that the homogeneous ordered solution is linearly unstable near
onset, and that solitary wave structures akin to the traveling bands,
arise at the nonlinear level.

Even though metric interaction zones are certainly of value in cases such as
shaken granular media \cite{Ramaswamy, Chate2010} and motility assays
\cite{Schaller2010,Sumino2012} where alignment arises mostly from 
inelastic collisions,
it has been argued recently \cite{Ballerini2008,Moussaid,Gautrais2012} that they
are not realistic in the context of higher organisms such as
birds, fish, or pedestrians, whose navigation decisions are likely to rely on 
interactions with neighbors defined using metric-free, ``topological'' criteria.
Statistical analysis of flocks of hundreds to a few thousand
individuals revealed that a typical starling interacts mostly with
its 7 or 8 closest neighbors, regardless of the flock density \cite{Ballerini2008}.
The realistic, data-based, model of pedestrian motion developed by 
Moussaid {\it et al.} relies on the ``angular perception landscape'' formed
by neighbors screening out others \cite{Moussaid}.

At a more theoretical level, the study of the Vicsek model with
Voronoi neighbors \cite{Ginelli2010b}
(those whose associated Voronoi cells form the first shell around the central cell)
has shown that metric-free interactions are relevant at the collective scale: 
in particular, the traveling bands mentioned above disappear,
leaving only a Toner-Tu-like phase. 
Below, we show that the introduction of Voronoi neighbors
suppresses the density-segregated phase
in other variants of the Vicsek model.
In spite of the recognized importance of metric-free interactions,
no continuous field equations describing the above models are available which 
would help put the above findings on firmer theoretical ground.

In this Letter, starting from Vicsek-style microscopic models with Voronoi neighbors,
we derive nonlinear field equations for active matter with metric-free interactions 
using a kinetic approach well-controlled near the onset of orientational order. 
We show that the density-independent collision rate per particle
characteristic of these systems suppresses the linear instability of the
homogeneous ordered phase and the nonlinear density segregation
in agreement with microscopic simulations.
We finally discuss the consequences of our findings for the relevance of 
metric-free interactions.

Let us first stress that with metric-free interactions, say with Voronoi neighbors,
the tenet of the Boltzmann equation approach \cite{Cercignani}, the assumption that the system is dilute enough
so that it is dominated by binary collisions, is never justified since, after all, 
a particle is almost constantly interacting with the same number of neighbors.
Here, instead, we introduce an interaction rate per unit time.
In the low rate limit, binary interactions dominate, and
one can proceed ``as usual''.

Our starting point is not a Vicsek-style model, 
but a microscopic rule in the same universality class:
$N$ point particles move at constant speed $v_0$ on a $L\times L$ torus; 
their heading $\theta$ is submitted to two different dynamical mechanisms, 
``self-diffusion'' and aligning binary ``collisions''. 
In self-diffusion, 
$\theta$ is changed into $\theta'=\theta+\eta$ with a probability $\lambda$ 
per unit time, where $\eta$ is a random variable drawn from a
symmetric distribution 
$P_\sigma(\eta)$ of variance $\sigma^2$. 
Aligning ``collisions'' occur at rate $\alpha$ per unit time with a 
{\it randomly chosen Voronoi neighbor}. Then $\theta$ is changed to
$\theta' = \Psi(\theta,\theta_{\rm n})+\eta$
where $\theta_{\rm n}$ is the heading of the chosen neighbor and the noise $\eta$ 
is also drawn, for simplicity, from $P_\sigma(\eta)$.
Isotropy is assumed,
namely $\Psi(\theta_1+\phi,\theta_2+\phi)=\Psi(\theta_1,\theta_2)+\phi \,[2\pi]$.
For the case of ferromagnetic alignment treated in detail below, 
$\Psi(\theta_1,\theta_2) \equiv \arg (e^{i\theta_1}+e^{i\theta_2})$.

The evolution of the one-particle phase-space distribution $f(\mathbf{r},\theta,t)$ 
(defined over some suitable coarse-grained scales) is governed by the
Boltzmann equation
\begin{equation}
\label{eqB}
\partial_t f({\bf r},\theta,t) + v_0\,{\bf e}(\theta) \cdot \nabla f({\bf r},\theta,t) =
I_{\rm diff}[f] +I_{\rm coll}[f]
\end{equation}
where ${\bf e}(\theta)$ is the unit vector along $\theta$.
The self-diffusion integral is
\begin{equation}
\label{eqdiff}
I_{\rm diff}[f]\!=\! -\lambda f(\theta) +\lambda\! \!\int_{-\pi}^{\pi}\!\!\! d\theta'\!\! \int_{-\infty}^{\infty}\!\!\! d\eta\, P_\sigma(\eta)\, 
\delta_{2\pi}(\theta'\!-\theta+\eta) f(\theta')
\end{equation}
where $\delta_{2\pi}$ is 
a generalized Dirac delta imposing 
that the argument is equal to zero modulo $2\pi$.
In the small-$\alpha$ limit, orientations are decorrelated 
between collisions
(``molecular chaos hypothesis''), and one can write:
\begin{eqnarray}
\label{eqcoll}
\!\!\!\!\!I_{\rm coll}[f]\!\!&=&\!-\alpha f(\theta)\!+\!\frac{\alpha}{\rho(\mathbf{r},t)} \!
\int_{-\pi}^{\pi}\! d\theta_1 \!\int_{-\pi}^{\pi}\! d\theta_2 \! \int_{-\infty}^{\infty}\!\! d\eta \nonumber\\
\!&\times& P_\sigma(\eta) f(\theta_1) f(\theta_2) \,
\delta_{2\pi}(\Psi(\theta_1,\theta_2)\!-\!\theta\!+\!\eta)
\end{eqnarray}
The main difference with the metric case treated in \cite{bertin}
is the ``collision kernel'', which is
independent from relative angles and inversely proportional to
the local density
\begin{equation}
\label{eqrho}
\rho(\mathbf{r},t) = \int_{-\pi}^{\pi} f(\mathbf{r},\theta,t)\, d\theta.
\end{equation}
Note that, in agreement with basic properties of models with metric-free interactions, 
Eq.~(\ref{eqB}), together with definitions (\ref{eqdiff}-\ref{eqrho}), is 
left unchanged by an arbitrary normalization of $f$ (and thus of $\rho$)
and thus does not depend on the global density $\rho_0=N/L^2$.
Furthermore, a rescaling of time and space allows to set 
$\lambda=v_0=1$, a normalization we adopt in the following.

Equations for hydrodynamic fields are obtained by expanding 
$f(\mathbf{r},\theta,t)$ in Fourier series, yielding the Fourier modes
$\hat{f}_k({\bf x}, t) = \int_{-\pi}^{\pi} d \theta 
f({\bf x}, \theta, t) e^{i k \theta}$
where $\hat{f}_{k}$ and $\hat{f}_{-k}$ are complex conjugates,
$\hat{f}_0=\rho$, and the real and imaginary parts of $\hat{f}_1$ are the 
coordinates of the momentum vector ${\bf w}=\rho\, {\bf P}$ with ${\bf P}$ 
the polar order parameter field. 
Using these Fourier modes, the Boltzmann equation (\ref{eqB}) 
yields an infinite hierarchy:
\begin{eqnarray}
\!\!\!\p_t \hat{f_k} + (\triangledown\hat{f}_{k-1}\!+\!\triangledown^*\! \hat{f}_{k+1})\!
& =& (\hat{P}_k-1-\alpha) \hat{f_k}\nonumber\\
& +& \frac{\alpha}{\rho} \hat{P}_k \sum_{q=-\infty}^\infty\! J_{kq}\hat{f}_q \hat{f}_{k-q}
\label{eq:kgen}
\end{eqnarray}
where the complex operators $\triangledown=\partial_x+i\partial_y$ and 
$\triangledown^*=\partial_x-i\partial_y$ have been used, 
the binary collision rate $\alpha$ is now expressed in the rescaled units,
$\hat{P}_k=\int_{-\infty}^{\infty} d\eta P_{\sigma}(\eta) e^{i k \eta}$
is the Fourier transform of $P_\sigma$,
and $J_{kq}$ is an integral depending on the alignment rule $\Psi$.
Below, we specialize to the case of ferromagnetic alignment, for which
\begin{equation}
J_{kq}= \frac{1}{2 \pi} \int_{-\pi}^{\pi} d \theta
\cos[(q-k/2)\theta] \;.
\end{equation}
For $k=0$ the r.h.s. of Eq.~(\ref{eq:kgen}) vanishes and 
one recovers the continuity equation:
\begin{equation}
\p_t \rho + \nabla \cdot {\bf w} = 0.
\label{eq:cont}
\end{equation}
To truncate and close this hierarchy, 
we assume the following scaling structure, valid 
near onset of polar order, assuming, 
in a Ginzburg-Landau-like approach, 
small and slow variations of fields:
\begin{equation}
\rho-\rho_0 \sim\epsilon, \;
\hat{f}_k  \sim \epsilon^{|k|}, \;\triangledown  \sim \epsilon, \; \partial_t \sim \epsilon
\end{equation}
Note that the scaling of space and time is in line with the propagative 
structure of our system \cite{Igor}.
The lowest order yielding non-trivial, well-behaved equations is $\epsilon^3$: 
keeping only terms up to this order, equations for $\hat{f}_{k>2}$ identically vanish,
while $\hat{f}_2$, being slaved to $\hat{f}_1$, 
allows to close the $\hat{f}_1$ equation which reads, 
in terms of the momentum field ${\bf w}$:
\bea \label{eq-hydro}
\p_t \w &+& \gamma (\w \cdot \nabla) \w =
-\frac{1}{2} \nabla \rho + \frac{\kappa}{2}\nabla\w^2\\
\nonumber
&+& (\mu-\xi \w^2) \w +\nu \nabla^2 \w -\kappa (\nabla \cdot \w)\w
\eea
Apart from some higher order terms we have discarded here, 
this equation has the same form as the one derived
in \cite{bertin} for metric interactions, but with different transport coefficients:
\begin{equation}
\begin{array}{ll}
\label{eq-coeff}
\mu = \left( \frac{4\alpha}{\pi} \!+\! 1 \right)  \hat{P}_1 \!-\! 
(1\!+\!\alpha)\;\;\; &
\nu = [4 (\alpha \!+\! 1 \!-\! \hat{P}_2)]^{-1} \\
\gamma = \nu\frac{4\alpha}{\rho}\! \left[\hat{P}_2 \!-\! \frac{2}{3\pi}\hat{P}_1 \right] \;\; &
\kappa = \nu  \frac{4\alpha}{\rho}\! \left[\hat{P}_2 \!+\! \frac{2}{3\pi}\hat{P}_1 \right] \\
\xi = \nu\! \left[\frac{4\alpha}{\rho}\right]^2\! \frac{1}{3\pi} \hat{P}_1\hat{P}_2 &
\end{array}
\end{equation}
Note first that, contrary to the metric case, the coefficient $\mu$ of the linear
term does {\it not} depend on the local density $\rho$; coefficients of the
non-linear terms depend on density to compensate the density dependence of $\w$.
Note further that $\nu$, $\kappa$, and $\xi$ are positive since $0<\hat{P}_k<1$, so that
in particular the nonlinear cubic term is always stabilizing.
For an easier discussion, we consider now the Gaussian distribution 
$P_\sigma(\eta)=\frac{1}{\sigma\sqrt{2\pi}} 
\exp[-\frac{\eta^2}{2\sigma^2}]$ for which
$\hat{P}_k=\exp[-k^2\sigma^2/2]$.
Then $\mu$ is negative for large $\sigma$ (where the trivial $\w=0$ solution
is stable with respect to linear perturbations), and changes sign for
\cite{NOTE6} 
\begin{equation}
\sigma_{\rm c}^2 = 2\ln\left( \frac{1+4\alpha/\pi}{1+\alpha} \right).
\end{equation}
For $\sigma<\sigma_{\rm c}$, the nontrivial homogeneous solution $\rho=\rho_0$,
$\w = \w_1 \equiv  \mathbf{e}\,\sqrt{\mu/\xi}$ 
(where $\mathbf{e}$ is an arbitrary unit vector) exists and is stable to 
homogeneous perturbations.

We now focus on the linear stability of $\w_1$ with respect to arbitrary
wavevector ${\bf q}$.
Because we want to discuss later differences between the metric and metric-free cases,
we keep a formal $\rho$-dependence of the linear transport
coefficient. Linearization around $\w_1$ yields
\bea \label{eq-lin}
\p_t \delta \rho &=& -  \nabla \cdot \delta \w  \\
\p_t  \delta \w \!&=&\! -\gamma (\w_1 \!\cdot\! \nabla)  \delta \w \!-\!
\frac{1}{2} \nabla  \delta \rho \!+\! \nu \nabla^2 \delta \w 
\!+\! \kappa \nabla (\w_1 \! \cdot \! \delta \w) \nonumber\\
\nonumber
&-&\!\kappa \w_1 (\nabla \! \cdot \!  \delta \w )  
\!-\! 2 \xi \w_1 (\w_1 \! \cdot \! \delta \w) \!+\! \left(\mu' \!-\!\xi' \w_1^2 \right)\w_1 \delta \rho
\eea
where primes indicate derivation with respect to $\rho$.
%
Using the ansatz 
$(\delta \rho ({\bf x}, t),\delta \w ({\bf x}, t))=
\exp(s t + i {\bf q} \cdot {\bf x}) (\delta \rho_{\bf q}, \delta \w_{\bf q})$ 
allows to recast Eq. (\ref{eq-lin}) as an eigenvalue problem for $s$.

\begin{figure}
\centering
\includegraphics[draft=false,clip=true,width=0.5\textwidth]{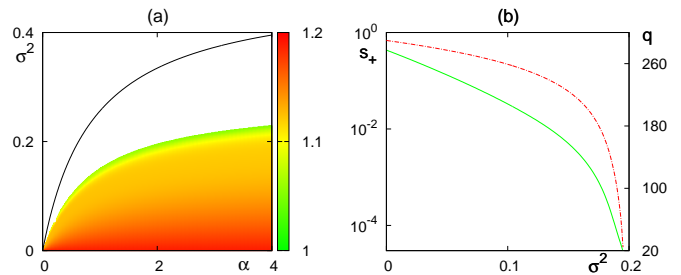}
\caption{(color online) Continuous theory for self-propelled particles aligning
ferromagnetically (Eqs.~(\ref{eq:cont}) and (\ref{eq-hydro}) 
with coefficients (\ref{eq-coeff}), Gaussian noise). 
(a) Phase diagram in the $(\alpha,\sigma^2)$ plane. The solid line marks the 
order-disorder transition. The homogeneous ordered solution $\w_1$
exists below this line and is linearly stable above the coloured linear 
instability region. The colour scale (in radians) 
indicates the most unstable wavevector direction 
$\phi$. (b) Modulus $q$ of the most unstable wavevector (green line)
and the real part of its corresponding eigenvalue $s_+$ (dashed red line) as a function of $\sigma^2$ at $\alpha=2$.
} 
\label{fig1} 
\end{figure}

We have solved numerically this cubic problem for the 
metric-free case using coefficients
(\ref{eq-coeff}) and Gaussian noise in the full $(\alpha,\sigma)$ parameter plane.
The resulting stability diagram, presented in Fig.~\ref{fig1}, shows that, 
contrary to the metric case, the homogeneous ordered phase is stable near onset. 
Like in the metric case \cite{NOTE5},
there exists an instability region to oblique wavevectors of large
modulus rather far from the transition line. Given that microscopic simulations
show no sign of similar instabilities,
we believe that the existence of this region,
situated away from the validity domain of our approximations, is an artifact
of our truncation ansatz.
 
At the nonlinear level, we performed numerical simulations
\cite{Numerics} of Eq.~(\ref{eq-hydro}) (again
with coefficients (\ref{eq-coeff}) and Gaussian noise) starting from
initial conditions with large variations of both $\rho$ and $\w$. With parameters
$\alpha$ and $\sigma^2$ in the ordered stable region of Fig.~\ref{fig1}, 
we always observed relaxation towards the linearly-stable homogeneous solution $\w_1$,
albeit after typically long transients. Starting in the unstable region,
the solution blows up in finite-time, signaling that indeed our equation is
ill-behaved when considered too far away from onset.

The stabilization of the near-threshold region by metric-free interactions
can be directly traced back to the absence of $\rho$ dependence of $\mu$, 
in agreement with remarks in \cite{bertin, Marchetti2}
where the long-wavelength instability
of $\w_1$ was linked to $\mu'>0$. 
In the long wavelength limit $q=|{\bf q}| \ll 1$, 
the eigenvalue problem can be solved analytically with relative ease. 
The growth rate $s$ is the solution of the cubic equation
\begin{equation}
s^3+\beta_2 s^2 + \beta_1 s + \beta_0 = 0
\label{cubic}
\end{equation}
where the coefficients, to lowest orders in $q$, are given by
\bea 
\label{eq-beta}
\beta_2&=& 2 \mu + 2 i q \sqrt{\frac{\mu}{\xi}} \gamma \cos\phi +
2 q^2 \nu \nonumber\\
\beta_1&=& i q  \sqrt{\frac{\mu}{\xi}} \cos\phi \left[2 \gamma
  \mu + \left(\mu'-\mu\frac{\xi'}{\xi} \right)\right]\nonumber\\
 &+& q^2 \left[ 2 \mu \nu + \frac{1}{2} - \frac{\mu}{\xi} \left(\gamma^2 \cos^2\phi + \kappa^2 \sin^2\phi\right)\right]\\
\beta_0&=&  \mu q^2 \left[\left(\frac{\mu \xi'}{\xi^2}\!-\!\frac{\mu'}{\xi}\right)\left(\gamma \cos^2\phi \!+\! \kappa \sin^2\phi\right)+  \sin^2\phi\right]
\nonumber\\
&+&i q^3 \sqrt{\frac{\mu}{\xi}}\cos \phi \left[\left(\mu'-\mu\frac{\xi'}{\xi} \right)\nu+ \frac{\gamma}{2}\right] \nonumber
\eea
and $\phi$ is the angle between ${\bf q}$ and $\w_1$
(which has been chosen parallel to the abscissa). 
Near threshold, where our truncation is legitimate and $\mu \sim \epsilon^2$, 
two eigenvalues are always stable and linear stability is controlled by the dominant solution real part
\begin{equation}
s_+ \approx \frac{q^2 \cos^2\phi}{8\mu} \left[\frac{(\mu')^2}{\mu\xi}-h(\phi) \right]
+ \mathcal{O}\left(\frac{q^2 \mu'}{\mu}\right)
\label{solution}
\end{equation}
with $h(0)=2$ and $h(\phi)=1$ otherwise \cite{NOTE4}.
This expression immediately shows that $s_+<0$ in the metric-free case where $\mu'=0$,
confirming the stability of the homogeneous ordered solution $\w_1$.
This stabilizing effect can be ultimately traced back to the (negative) pressure term
$-\nabla \rho $ appearing in Eq. (\ref{eq-hydro}).
Conversely, in the metric case for which $\mu\sim 0^+$ near threshold, $s_+$ is always
positive, yielding the generic long wavelength instability leading to density 
segregation in metric models.

To sum up at this point:
Through a kinetic approach, we have derived a continuous, ``hydrodynamic'' theory
for self-propelled particles aligning with topological neighbors
(``metric-free interactions''). 
Calculations were presented in full for the case
of ferromagnetic alignment, for which
we showed that the generic long-wavelength instability
of the homogeneous ordered solution present in metric models near onset is suppressed.
At the nonlinear level, we observed numerically that 
the density segregated solutions of metric models vanish, so that
the homogenous solution seems to be a global attractor.

These results rest on the independence of the linear coefficient $\mu$  
on $\rho$, a direct consequence of the
fact that the interaction rate per particle in topological models is
fixed only by geometrical constraints and does not grow with local
density. This property actually holds for any metric-free system:
all the linear coefficients $\mu_k$ 
appearing in the equations for $\hat{f}_k$ read
\be
\mu_k = \hat{P}_k - 1 -\alpha +\alpha \hat{P}_k \left( J_{kk} + J_{k0}\right)
\label{mu:gen}
\ee
and are thus independent of $\rho$.

\begin{figure}
\centering
\includegraphics[draft=false,clip=true,width=\columnwidth]{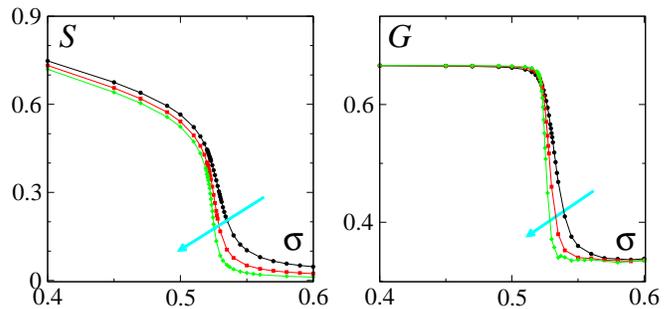}
\caption{(color online) Vicsek-style model with nematic alignment and topological
 neighbors, where $N=L^2$ particles move at speed $v_0=\frac{1}{2}$ 
on a $L\times L$ torus.
Headings and positions are updated at discrete timesteps
according to
$\theta_j^{t+1} = \arg\left[ \sum_{k\sim j}{\rm sign}[\cos(\theta_k^t - 
\theta_j^t)] e^{i\theta_k^t}  +n_j^t \sigma \xi_j^t \right]$ and
${\bf r}_j^{t+1} = {\bf r}_j^t + v_0 \, {\bf e}(\theta_j^{t+1})$
where ${\bf e}(\theta)$ is the unit vector along $\theta$, 
the sum is over the $n_j^t$ Voronoi neighbors of particle $j$ (including $j$ itself), 
and $\xi_j^t$ is a random  unit vector in the complex plane.
Global nematic order parameter $S=\langle |\frac{1}{N}\sum_k e^{i2\theta_k^t}|\rangle_t$ (a) and its Binder cumulant $G$ (b) vs $\sigma$ 
for $L=32$, 64, 128 (the arrows indicate increasing sizes), 
revealing a continuous transition.} 
\label{fig2} 
\end{figure}

Our other general conclusions also extend to the other basic classes of simple,
Vicsek-style collective motion: we have in particular worked out the case 
of polar particles with nematic alignment (``self-propelled rods'') for which 
$\Psi (\theta_1,\theta_2)=\arg (e^{i\theta_1}+{\rm sign}[\cos(\theta_1 - 
\theta_2)] \,e^{i\theta_2})$. While full details will be given in \cite{TBP},
we only sketch here the salient points. Let us first recall that 
with nematic alignment, the metric model studied at the microscopic level 
in \cite{Ginelli2010}, shows global nematic order.
In a large region of parameter space bordering onset,
order is segregated to a high-density stationary band oriented along it.
We have studied numerically the metric-free version of that
Vicsek-style model with 
Voronoi neighbors. Like for its ferromagnetic counterpart, no segregation
in bands is observed anymore, and the transition to nematic order is then
continuous (Fig.~\ref{fig2}).
These properties are well-captured, both in the metric and metric-free case, 
by a controlled hydrodynamic approach of the type presented here \cite{TBP}. 
The nematic symmetry of the problem requires to consider 
three hydrodynamic fields \cite{MCM2},
corresponding to the modes $k=0,1,2$ in Eq.~(\ref{eq:kgen}), with $\hat{f}_2$
coding for the nematic tensor field $\rho\,{\bf Q}$.
We have performed the analysis of the $5\times 5$ linear problem
expressing the stability of the 
homogeneous nematically-ordered solution ($\w=0$, $\rho\,{\bf Q}={\rm Cst.}$)
appearing at onset in both the 
metric and non-metric cases \cite{TBP}. 
Whereas the metric case shows a long wavelength,
transversal instability of the homogeneous ordered solution near onset, 
this solution is linearly stable in the metric-free case. Again, this difference
can be traced back to the $\rho$-dependence of the linear coefficients $\mu_k$.
At the nonlinear level, simulations indicate that the homogeneous ordered solution
is a global attractor in the metric-free case.

Our analysis can also be extended to
``non-ballistic'' active matter such as the driven granular rods model 
(``active nematics'') studied in \cite{Chate2006, Ramaswamy2} which, 
for metric interactions,
also shows near-threshold phase segregation \cite{Chate2006, Shi-Ma}. 
Simulations of the metric-free microscopic version (with Voronoi neighbors)
show no such segregation. 
In a kinetic approach, because active nematic particles move 
by non-equilibrium diffusive currents rather than by ballistic motion, 
the Boltzmann equation has to be replaced by
a more general master equation. But it is nevertheless possible to derive
a continuous theory which, in the metric-free case, yields a homogeneous ordered
phase stable near onset for essentially the same reasons as in the cases 
presented above \cite{TBP2}.

In conclusion, simple, Vicsek-style, models of active matter where self-propelled
particles interact with neighbors defined via non-metric rules 
({\it e.g.} Voronoi neighbors) are amenable, like their ``metric'' counterparts, 
to the construction of continuous hydrodynamic theories well-controlled near onset.
The relatively simple framework of Vicsek-style models offers a 
two-dimensional parameter plane 
which can be studied entirely. More complicated microscopic starting points, 
for instance where positional diffusion would also be considered, inevitably 
raise the dimensionality of parameter space.
We have shown here that the non-metric theories differ 
essentially from the metric ones in the independence of their linear coefficients
$\mu_k$ on the local density, a property directly linked to the fact that 
the collision rate per particle is constant in metric-free systems.
We have shown further that the homogeneous ordered phase is linearly stable 
near onset for metric-free systems, 
in contrast with the long-wavelength instability present in metric cases. 

We finally discuss the relevance 
(say in the renormalization-group sense) of metric-free interactions
in deciding active matter universality classes.
Our work has shown that the deterministic continuous theories of metric-free active 
matter systems are formally equivalent to those of their metric counterparts,
except for the density-dependence of the linear coefficients. This could be
taken as an indication, in the case of ferromagnetically-aligning particles, 
that the homogeneous, ordered, fluctuating phase observed
in the Vicsek model with Voronoi neighbors does {\it not} differ from
the Toner-Tu phase of its metric counterpart, in contradiction with
the numerical discrepancies between the two cases reported in \cite{Ginelli2010b}
about the scaling exponent of the anomalously-strong density fluctuations.
This calls for more extensive microscopic simulations 
assessing finite-size effects,
but also for incorporating effective noise terms, properly-derived
 in both cases, and to study the resulting 
field theories, a task left for future work.

\begin{acknowledgments}
We are grateful to I.S. Aranson, I. Giardina,
M.C. Marchetti, J. Prost, and J. Toner for enlightening discussions.
This work was initiated in the lively atmosphere of the Max
Planck Institute for the Physics of Complex Systems in Dresden, Germany,
within the Advanced Study Group 2011/2012:
Statistical Physics of Collective Motion.
\end{acknowledgments}


\begin{thebibliography}{99}
\bibitem{NOTE6}
While we are considering the small interaction rate limit, $\alpha$ is not necessarily
smaller than $1$ in the units chosen, for which $v_0=\lambda=1$.
\bibitem{NOTE5} In the metric case, this region of instability 
was overlooked in \cite{bertin}.
\bibitem{Numerics} We used a pseudospectral code with Euler
  timestepping ($\Delta t = 0.01$), with at least $256\times 256$ Fourier
  modes on square domains of linear size $L=64$ and anti-aliasing truncation.
\bibitem{NOTE4} For $\phi=\pm\pi/2$, one has to resort to higher orders, 
and then finds $s_+ \approx q^2 (-(\mu'/\mu)\kappa\xi+ 2\kappa^2 \xi - 4 \nu \xi^2 + \kappa \xi')/(8 \xi^2)$. This quantity is always negative for coefficients (\ref{eq-coeff}).
\end{thebibliography}
\end{document}